\documentclass[aps,preprint,nopacs]{revtex4-1}
\usepackage{graphicx}
\begin{document}
\bibliographystyle{apsrev}
\title{Influence of an electron-beam exposure on the surface plasmon resonance of gold nanoparticles}

\author{Mingxia Song$^{1,2}$ and G\'erard Colas-des-Francs$^{1}$ and Alexandre Bouhelier$^{1,*}$}

\affiliation{$^{1}$ Laboratoire Interdisciplinaire Carnot de Bourgogne, CNRS UMR 6303, Universit´e de Bourgogne,
9 Avenue Alain Savary, Dijon, France \\
$^{2}$ School of Environmental Science and Engineering, Nanjing University of information Science and Technology, Nanjing, China \\}

\email{alexandre.bouhelier@u-bourgogne.fr}


\begin{abstract}
Electron beam imaging is a common technique used for characterizing the morphology of plasmonic nanostructures. During the imaging process, the electron beam interacts with traces of organic material in the chamber and produces a well-know layer of amorphous carbon over the specimen under investigation. In this paper, we investigate the effect of this carbon adsorbate on the spectral 
position of the surface plasmon in individual gold nanoparticles as a function of electron exposure dose. We find an optimum dose for which the plasmonic response of the nanoparticle is not affected by the imaging process. The final publication is available at link.springer.com
\end{abstract}
 \maketitle
\section{Introduction}
\label{intro}

Sensing based upon the spectral wandering of localized surface plasmon resonances supported by metal nanoparticles is a major field of research and development~\cite{Stewart08,Hafnerreview,HalasReview}. By a synthetic engineering of the plasmon field, the sensitivity has now reached the detection level  of a single molecule and a unique binding event~\cite{quidantACS09,Hafner10,Zijlstra12,Sonnichsen12}. The development of sensitive plasmonic assays and the understanding of their optical responses was largely fostered by a precise knowledge of the morphology of the sensor. In this respect, electron microscopy is a routine technique for measuring interacting distances, determining shape anisotropy or simply observing the results of a synthesis~\cite{Hartland08,Liu2011,Margueritat,Halas12,Rogach}. 
However, when the accelerated electrons hit the specimen under observation, 
they not only produce secondary electrons pertaining information about the 
sample, but also interacts with carbon-containing residues inevitably present on the sample surface and in the vacuum chamber~\cite{Isabell99}. Depending on the imaging condition, this interaction produces an hydrocarbon buildup covering the surface of the sample
being imaged. Although this carbon uptake can be carefully controlled 
to fabricate plamonic structures~\cite{quidant10}, nanotips for field emission~\cite{Schiffmann93,Antognozzi97,Johnson02} 
or to serve as a mask for negative resist processes~\cite{Djenizian03}, it gradually reduces
the imaging contrast and causes spectroscopic artefacts~\cite{Griffiths10}. 
The deposition of such carbonaceous film during imaging process is evidently of great concern for plasmonic sensors because of their extreme sensitivity to surface adsorbate~\cite{quidantACS09,Hafner10,Zijlstra12,Sonnichsen12}. Strategies to alleviate film growth requires carbon volatilization in environmental SEMs~\cite{toth09} or usage of oxygen plasma etching device~\cite{Isabell99,Ebisawa,Vane12}. Their efficiency however depend on operating conditions and may not always mitigate the effect of an electron-beam induced carbon deposition.

In this paper, we evaluate the effect of a carbon contamination consequent of an electron-beam exposure on the surface plasmon resonance of individual gold nanoparticles. We find that after irradiation of the nanoparticles the position of the plasmon resonance red-shifts for modest exposure doses before undergoing a surprising blue-shift for larger irradiation doses. We then define an optimum condition for which the spectral position of the surface plasmon resonance before and after electron beam exposure remains the same. 

\section{Sample description and measurement protocol}
\label{sample}

\begin{figure}
\begin{center}
  \includegraphics[width=1\textwidth]{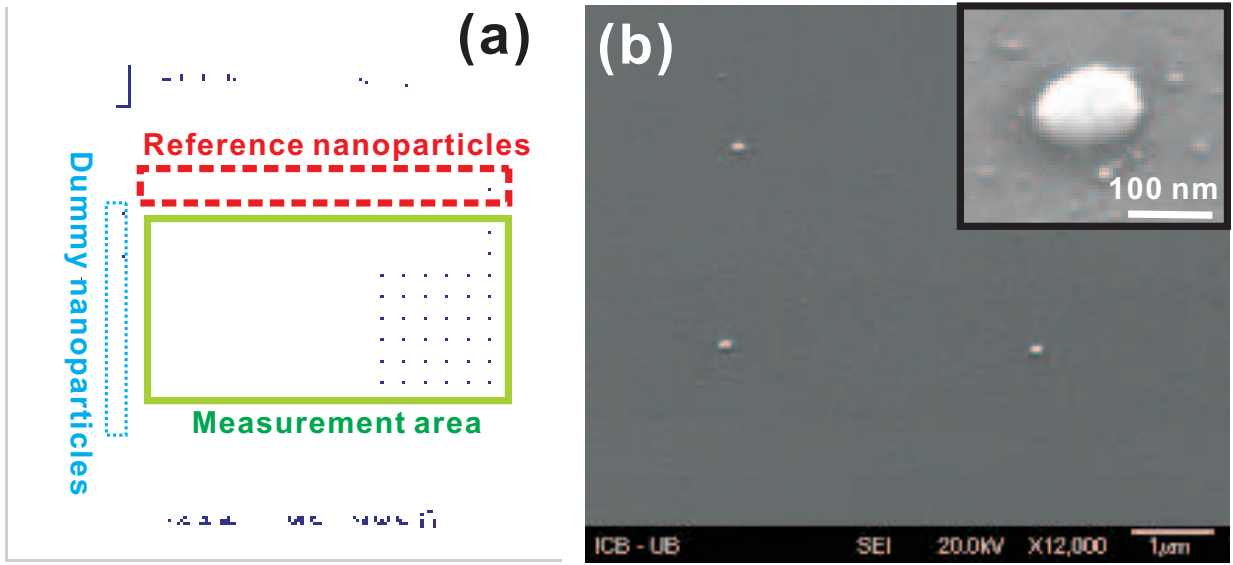}
  \end{center}
\caption{(a) Layout of the nanoparticles array. The first
nanoparticle on the left of each line is used as a reference
nanoparticle for determining the plasmonic response of pristine structures. The other nanoparticles are exposed to the electron beam with varying doses. The line in the blue box are dummy nanoparticles used to extract geometrical parameters. (b) Scanning electron micrograph of  Au nanoparticles in the array. Inset:
close up image of a Au nanoparticle.}
  \label{figure1}
\end{figure}

Gold nanoparticles are fabricated using 
standard electron electron beam lithography and liftoff process using a field-emission gun scanning electron microscope (JEOL 6500) operating with an acceleration voltage of 20~kV. The substrates used here are a 170~$\mu$m thick cover glass coated with a thin indium-tin-oxide (ITO) layer. The ITO coating is a thin transparent conductive layer (30~nm) used to evacuate the charge during the electron-beam lithography as well as during the electron-beam measurements discussed below. The layout of the gold nanoparticle arrays investigated is drawn in 
Fig.~\ref{figure1}(a). The array is constituted of 13 rows each composed of 9 nanoparticles. The nanoparticles of the top line (red dashed box) are reference nanoparticles that will not be exposed to the electron
beam. The unlabeled particles on the left (blue box) are dummy
units used to extract geometrical parameters. The spacing between nanoparticles in the measurement area (green rectangle) is 4~$\mu$m
while the pitch between lines is also 4~$\mu$m. Three rows are purposely missing in the array to ease nanoparticle recognition during the different measurement steps£. A scanning electron micrograph of typical Au nanoparticles in the array is depicted in Fig.~\ref{figure1}(b). The image was taken with a 45$^{\circ}$ tilt of the sample with respect to the electron beam. A
close up view of one nanoparticle is shown in the inset of
Fig.~\ref{figure1}(b). All the nanoparticles in the array have nominally a diameter of $\sim$125~nm and a thickness of $\sim$50~nm.

In this study, we used single-particle dark field spectroscopy to interrogate the spectral position of the surface plasmon resonance~\cite{mock02,bouhelier05jpcb}. Light from a 100~W tungsten lamp is focused on the array by a dark field condenser with a numerical aperture (N.A.) comprised between 0.8 and 0.95. The scattered light from the nanoparticles is
collected with an objective (40$\times$, N.A.=0.65) and is
sent to an imaging spectrometer through a 4-f imaging system.

The measurement protocol is as follow: First the spectrum of each nanoparticle in the array is measured before exposing the nanoparticles to the electron beam. With the interparticle distance considered here and the magnification of the collection objective, each nanoparticles are resolved by the imaging spectrometer and do not interfere with each other. By reducing the size of the slit on the spectrograph, individual spectra from a complete row of nanoparticles can be measured simultaneously. Once the spectral responses of the pristine nanoparticles are measured, the sample is transferred to the SEM and the nanoparticles placed in the measurement area (green box) are exposed
to the electron beam with varying irradiation doses $d$. All the nanoparticles on a given row in the measurement area are exposed to the electron beam under the same condition. The exposure dose for each row is illustrated in Table~\ref{table:Expose}. The current carried by the electron beam is evaluated by a pico-Amperemeter connected to a Faraday cage. $d$ is varied either at constant current or at constant exposure time. 
\begin{table}[ht]
\centering
\begin{tabular}{c c c c}
\hline
 Line & Current (nA) & Time (s) & Dose (nA.s)\\ \hline
1 & 0.885 & 5 & 4.4\\
2 & 0.885 & 10 & 8.8\\
3 & 0.885 & 15 & 13.2\\
4 & 0.885 & 20 & 17.7\\  \hline
5 & 0.229 & 5 & 1.1\\
6 & 0.229 & 10 & 2.3\\
7 & 0.229 & 15 & 3.4\\
8 & 0.229 & 20 & 4.6\\ \hline
9 & 0.089 & 5 & 0.4\\
10 & 0.089 & 10 & 0.9\\
11 & 0.089 & 15 & 1.3\\
12 & 0.089 & 20 & 1.8\\ \hline
\end{tabular}
\caption{Exposure doses used to irradiate the nanoparticles: dose increases either at constant
current and increasing exposure time, or by changing the current of
the electron beam and keeping the exposure time fixed.}
\label{table:Expose}
\end{table}

To quantify the carbon deposition upon electron irradiation, we also park the electron beam in a region at the bottom of the measurement area and irradiate the substrate surface with the same dose used to irradiate the adjacent nanoparticles.
After electron beam exposure,  the scattered response of the nanoparticles is measured again and is compared to the original spectrum. The spectra of the reference nanoparticles (red box in Fig.~\ref{figure1}) are also measured again to confirm that the references remain the same before and after conducting the contamination process within the measurement area.

\section{Dark-field spectroscopy}
\label{spectroscopy}

\begin{figure}
\begin{center}
  \includegraphics[width=1\textwidth]{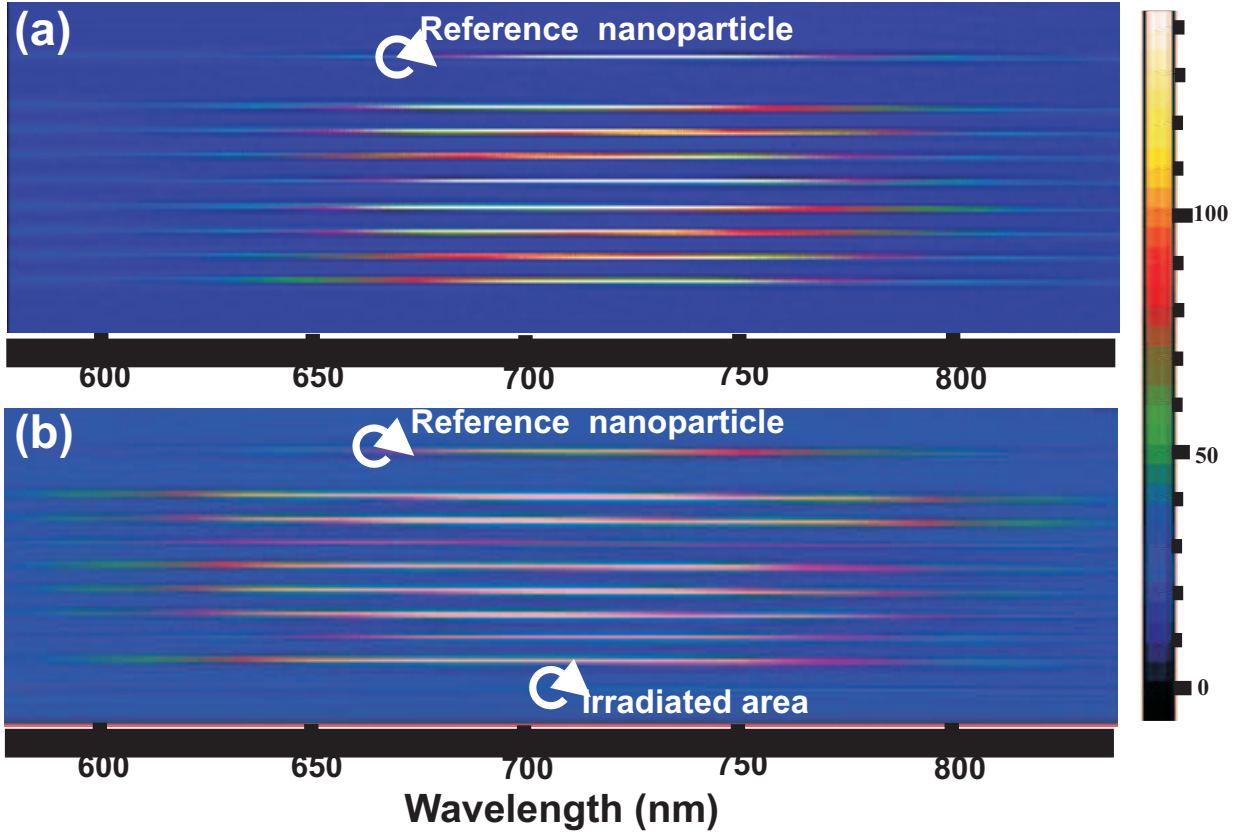}
  \end{center}
\caption{Hyperspectral images from a row of of 9 Au nanoparticles (a) before electron
exposure and (b) after exposure with a dose of 13.2~nA.s. The signals at the top of each images are scattered spectra from the reference nanoparticles. The additional spectrum at the bottom of (b) arises from an irradiated zone of the substrate exposed under the same electron dose.}
  \label{figure2}
\end{figure}

Figure~\ref{figure2} displays the background-corrected
hyperspectral images of 9 adjacent Au nanoparticles before and after
electron beam exposure, respectively. The exposure dose is $d$=13.2~nA.s. The first line at the top of both images is the spectrum of
the reference nanoparticle. Small variation in the intensity of the scattered light are observed between the individual nanoparticles and are attributed to inherent limitations of the fabrication procedure. The surface plasmon resonances are however fairly constant within the row.

\begin{figure}
\begin{center}
  \includegraphics[width=1\textwidth]{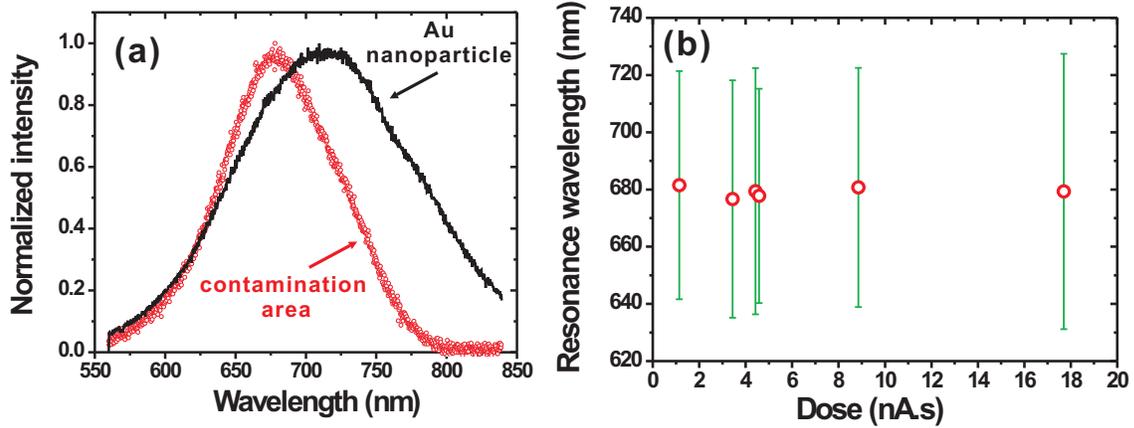}
  \end{center}
\caption{(a) Extracted spectra of a reference Au nanoparticle and that of contaminated
area with $d$=4.4~nA.s. The spectrum of the contaminated area strongly resembles that
of a surface plasmon resonance. (b) Evolution of the wavelength maximum scattered from contamination
carbon dot as a function of electron dose. The error bars represent the full-width at half-maximum of the resonance.}
  \label{figure3}
\end{figure}

Figure~\ref{figure2}(b) shows an additional scattering event from an entity at the bottom of the hyperspectral image. This signal originates from the surface contamination that is formed when the bare substrate is irradiated under the same dose setting as the nanoparticles.  Interestingly, the light scattered from this
area presents a resonance-like shape as shown in Fig.~\ref{figure3}(a).
The spectra resembles that of a plasmonic response of an
unexposed reference nanoparticle. We find that peak wavelength of the the contamination dots  does not significantly depend on $d$ as illustrated in Fig.~\ref{figure3}(b). The spectral position of the scattering maximum is constant with doses. The error bars represent the full-width at half maximum of the measured spectra.  Since the scattering response of the contamination dot is dispersive, the spectrum
measured from an irradiated nanoparticle must therefore include a linear combination of
the nanoparticle's response and that of the carbon residue.

\section{Effect of electron irradiation}

\begin{figure}
\begin{center}
  \includegraphics[width=1\textwidth]{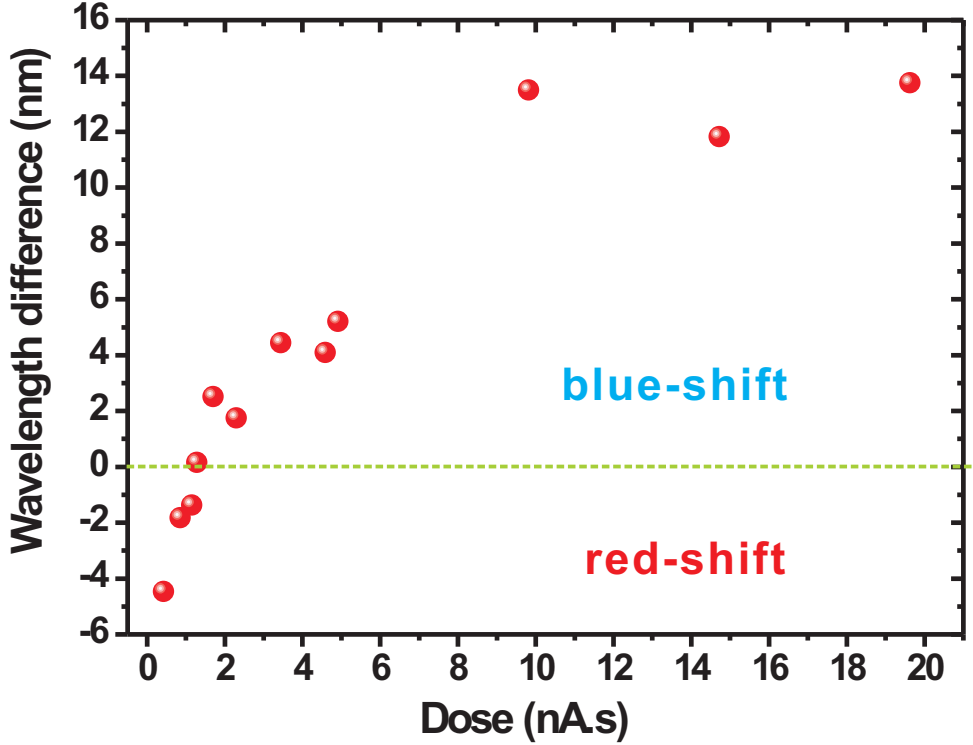}
  \end{center}
\caption{Wavelength difference $\delta_{\lambda}$ of the spectral maximum scattered from the
nanoparticles before and after electron exposure as a function of dose. A negative value  of $\delta_{\lambda}$ indicates a red-shift of the resonance compared to the reference spectrum and a positive value a blue-shift.}
  \label{figure4}
\end{figure}

The difference of the resonance wavelength for each Au nanoparticle
before and after electron irradiation is analyzed by performing a Gaussian
fitting procedure on each spectrum. The resonance wavelength of the
pristine particles $<\lambda_{\rm o}>$ is obtained by taking the average
spectra of the 9 nanoparticles forming the same row, including the
reference system. The resonance wavelength of the nanoparticles after
contamination $<\lambda_{\rm c}>$ is obtained by averaging the scattered peaks of
the 8 nanoparticles that were exposed to the electron beam. The
wavelength difference $\delta_{\lambda}$ 
before and after electron contamination is thus defined as $\delta_{\lambda} =
<\lambda_{\rm o}> - <\lambda_{\rm c}>$. The results are plotted as a function
of exposure dose in Fig.~\ref{figure4}. $\delta_{\lambda}<0$
indicates a redshift of the resonance after exposure compared to the reference spectrum while a positive
$\delta_{\lambda}$ signifies a blue-shift of the response. For the smallest exposure dose, the
resonance wavelength of Au nanoparticles has redshifted compared to the
reference nanoparticle as expected from the adhesion of surface
adsorbates~\cite{Kelly03,mcfarland03}. For increasing values of $d$, we however systematically observe a shift of the resonances toward the blue part of the spectrum. After $d$=2~nA.s, the spectral maximum is even blue-shifted compared to the reference spectrum as indicated by the positive value of $\delta_{\lambda}$. This blue-shift steadily increases before saturating at $\sim$13~nm for doses higher than 10~nA.s.

\section{Discussion}
\begin{figure}
\begin{center}
  \includegraphics[width=1\textwidth]{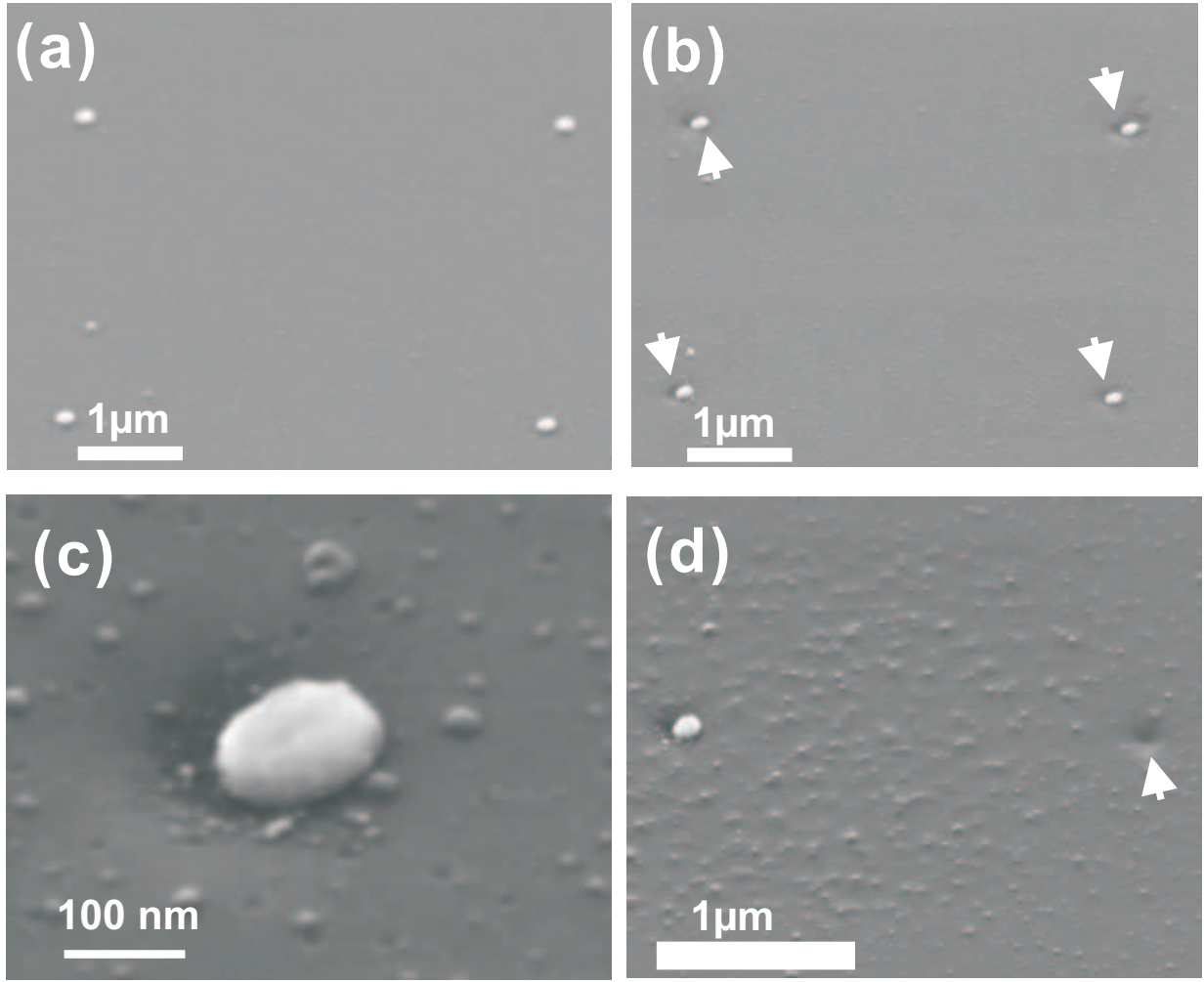}
  \end{center}
\caption{SEM images of four Au pristine nanoparticles (a) and after
electron exposure (b). The arrows indicates the surface deformation resulting from the electron beam.  (c) Close-up image of one Au particle
after irradiation showing a hollow surrounding the nanoparticle. (d) SEM image of an irradiated Au nanoparticle together with a pit on the surface (arrow) formed on
its right side under the same exposure dose.}
  \label{figure5}
\end{figure}

In order to explain the evolution of $\delta_{\lambda}$ with $d$, we fabricated an array of Au nanoparticles and
imaged  the nanoparticles before and after electron-beam irradiation by SEM without conducting the spectroscopy of section~\ref{spectroscopy}. We show in Fig.~\ref{figure5} (a) and (b) SEM images of four Au nanoparticles taken before and after electron beam exposure. The lower pair in Fig.~\ref{figure5} (b) is irradiated with a dose $d$=13.2~nA.s and the upper pair with $d$= 17.7~nA.s.  Under these doses, the SEM images reveal the appearance of small depressions surrounding the nanoparticles as indicated by the arrows in Fig.~\ref{figure5} (b). The size and depth of the depressions increase with $d$ as indicated by the contrast difference of the hollows around the lower and upper pairs. Deformation of the surface resulting from the impact of an electron beam has been reported for various class of materials ranging from III-V compounds, oxides and polymers~\cite{Ogawa99,storm05,akiba10} and is generally understood as a strain relaxation of the surface irradiated. A close up view of one
irradiated Au nanoparticle with a depression around is shown in Fig.~\ref{figure5}(c). In
Fig.~\ref{figure5}(d), the depression site formed by parking the electron beam on the bare substrate is readily observed $\sim$
2.5$\mu$m at the right of the nanoparticle (arrow). At this location, the
substrate was exposed under the electron beam with the same dose as on the nanoparticle on the left. It is clear that under a high exposure dose, the carbon contamination induced by the electron beam is etched away~\cite{knowles07} and  the impact of the electron beam leads to a morphological change of the substrate surface. We thus study the cross-over dose for which the surface plasmon response of the Au nanoparticles is affected by a carbonization of its surface or by a local change of the substrate topology.

\begin{figure}
\begin{center}
  \includegraphics[width=1\textwidth]{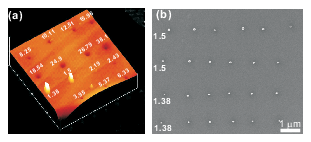}
  \end{center}
\caption{(a) 3D AFM image of carbon contamination 
tips and depressions. The exposure doses used are labeled next to the surface features. Transition from the growth of a dot to the formation of hollow occurs between 2.19 nA.s and 2.43 nA.s. The scanned area is a square of 12$\mu$m$\times$12$\mu$m. (b) SEM image of contamination dots deposited on ITO 
with exposure doses 1.38 nA.s and 1.5 nA.s demonstrating the robustness of the process.}
  \label{figure6}
\end{figure}

Figure~\ref{figure6}(a) shows a three-dimensional rendering of an atomic force microscope (AFM) image of a 12$\mu$m $\times$ 12$\mu$m area of a pristine ITO-covered substrate. The zone was exposed locally to the electron beam 
using an array of doses. $d$ was
changed from 1.38~nA.s to 38.1~nA.s. The exposure dose is marked on the
top of each parking position. For the lowest exposure
dose ($d$=1.38~nA.s and 1.5~nA.s) carbon
tips are formed (third line from the top). The growth of these protrusions are fairly reproducible as shown by the array depicted in the SEM image of Fig.~\ref{figure6}(b). Here a series of carbon contamination dots
formed with $d$=1.38~nA.s and $d$=1.5~nA.s emphasizes the robustness of the
contamination. The carbon dots formed with an exposure dose of 1.5~nA.s
are slightly larger than those formed at $d$=1.38~nA.s.

\begin{figure}
\begin{center}
  \includegraphics[width=1\textwidth]{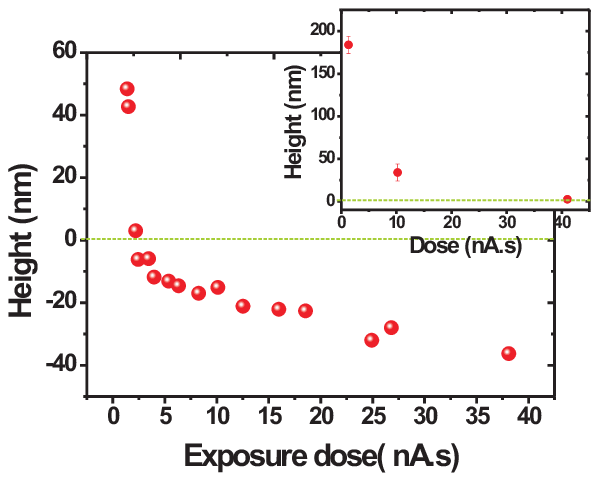}
  \end{center}
\caption{Height of the carbon dots and the depressions formed by using 
different exposure dose on the surface of ITO. Inset: results obtained from a Raith Pionner microscope. }
  \label{figure7}
\end{figure}

When $d$ increases to 2.19~nA.s, a small
protrusion remains visible in the AFM image of Fig.~\ref{figure6}(a).  At the next dose, however ($d$=2.43~nA.s)  a depression is replacing the contamination dot as indicated by the contrast reversal of the image.
Figure~\ref{figure7} displays the height of the carbon tips and the
depressions extracted from Fig.~\ref{figure6}(a). As the exposure dose increases,
the height of the contamination dots decreases and starts to form depressions at around $d$=2.5~nA.s.  Within the experimental errors, this dose threshold coincides  approximately to the dose for which $\delta_{\lambda}$=0 (Fig.~\ref{figure4}). For higher value of $d$, the depression becomes deeper as the exposure dose increases and tends to saturate after 10~nA.s, a value consistent with the saturation of $\delta_{\lambda}$ in Fig.~\ref{figure4}. Transition from carbon growth on the exposed area to a volatilization of the residues originates from the relative efficiencies of the deposition and etch processes~\cite{knowles07}.  Growth is more efficient but saturates at high beam current to the limited arrival of hydrocarbon species. Conversely, the etching process is less efficient but benefits from the large supply of water molecules present on the substrate and in the chamber.

From the above considerations, we can then conclude that when the
exposure dose is small, carbon contamination tips formed on the top
surface of the Au particle introduce a small but measurable
red-shift of the plasmon resonance. The redshift results from a modified
electrical environment attributed to the hydrocarbon adsorbate.
As the exposure dose increases, the electron beam is no
longer depositing carbon residues and contributes to the desorption
of the contamination. At approximately 2.5~nA.s, the position of the
resonance shows that the nanoparticle before and after electron-beam
exposure is the same. This is confirmed by the vanishing height of the contamination dot. Hence, this exposure dose defines the optimal imaging
condition for observing plasmonic nanoparticles without altering
their spectroscopic properties. For larger exposure, the geometry of
the interface is changed to a concave surface. The blue-shift of the
resonance is therefore understood as joint effects arising from a linear combination of
the scattering response of the indentation convoluted with that of
the nanoparticle, and a modified interface from a planar surface to a
concave geometry. We ruled out an  irreversible change of the nanoparticle morphology resulting from the repeated impact of the electrons to explain the wavelength shift of the resonance. Under the operating condition of the SEM, the electron-induced  temperature rise remains far below the melting temperature of the nanoparticle~\cite{Reimer,Borel76} and sputtering of metals is unlikely in a 20 KV SEM~\cite{Egerton}. 

We verified that the general trend observed in Fig.~\ref{figure7} is not typical from the instrument used in this study and can be generalized to other SEMs. To this purpose we measured the height of carbon dots on a pristine ITO with a Raith microscope (Pioneer) operating at 20~KV under three different doses. Results are shown in the inset of Fig.~\ref{figure7}. The trend is qualitatively reproduced with this microscope, however, carbon growth is inhibited for a larger dose ($\sim$ 40~nA.s) compared to Fig.~\ref{figure7}. While the overall growth and etch processes of the carbon residues are consistent between instruments, the specific dose for which the surface plasmon resonance is unaffected by the electron flux needs to be predetermined for each SEM.

\section{Conclusion}
We have studied the effect of an electron-beam exposure on the response of plasmonic
nanoparticles upon imaging with a SEM. The spectra obtained for Au nanoparticles before and after
irradiation indicate the presence of carbon contamination on the
surface of the nanoparticle for smaller exposure doses.
When the exposure dose increases, AFM and SEM imaging demonstrate
that the carbon deposits are etch away and the substrate around Au particles is deformed by the
electron beam to form a depression. Under these exposure conditions,
the surface plasmon resonance of the nanoparticle systematically
blue-shifts passing by its original value before electron beam exposure. An optimum dose can therefore be determined to safely image
plasmonic nanostructures with SEM without altering their spectroscopic response.

\section{Acknowledgements}
The research leading to these results has received funding from the European Research Council under the European Community's Seventh Framework Program FP7/2007-2013 Grant Agreement no 306772. This project is in cooperation with the regional council of Burgundy under PARI SMT3 and the Labex ACTION program (contract ANR-11-LABX-01-01).  M. S. acknowledges the support of the Jiangsu Specially-Appointed Professor Program (R2012T01) and the Foundation for special project on the Integration of Industry, Education and Research of Jiangsu Province (BY2012028).

\bibliographystyle{spphys}       

\begin{thebibliography}{99}
\providecommand{\url}[1]{{#1}}
\providecommand{\urlprefix}{URL }
\expandafter\ifx\csname urlstyle\endcsname\relax
  \providecommand{\doi}[1]{DOI \discretionary{}{}{}#1}\else
  \providecommand{\doi}{DOI \discretionary{}{}{}\begingroup
  \urlstyle{rm}\Url}\fi

\bibitem{Stewart08}
M.E. Stewart, L.B. Anderton, Christopher R.and~Thompson, J.~Maria, S.K. Gray,
  J.A. Rogers, R.G. Nuzzo, Chem. Rev. \textbf{108}, 494 (2008)

\bibitem{Hafnerreview}
K.M. Mayer, J.H. Hafner, Chem. Rev. \textbf{111}(6), 3828 (2011)

\bibitem{HalasReview}
N.J. Halas, S.~Lal, W.S. Chang, S.~Link, P.~Nordlander, Chem. Rev.
  \textbf{111}(6), 3913 (2011)

\bibitem{quidantACS09}
S.S. Aćimović, M.P. Kreuzer, M.U. González, R.~Quidant, ACS Nano
  \textbf{3}(5), 1231 (2009)

\bibitem{Hafner10}
K.M. Mayer, F.~Hao, S.~Lee, P.~Nordlander, J.H. Hafner, Nanotech.
  \textbf{21}(25), 255503 (2010)

\bibitem{Zijlstra12}
P.~Zijlstra, M.~Paulo, Pedro M. R.and~Orrit, Nature Nanotech. \textbf{7},
  379 (2012)

\bibitem{Sonnichsen12}
I.~Ament, J.~Prasad, A.~Henkel, S.~Schmachtel, C.~S\"onnichsen, Nano Lett.
  \textbf{12}(2), 1092 (2012)

\bibitem{Hartland08}
M.~Hu, C.~Novo, A.~Funston, H.~Wang, H.~Staleva, S.~Zou, P.~Mulvaney, Y.~Xia,
  G.V. Hartland, J. Mater. Chem. \textbf{18}, 1949 (2008)

\bibitem{Liu2011}
N.~Liu, M.~Hentschel, T.~Weiss, A.P. Alivisatos, H.~Giessen, Science
  \textbf{332}(6036), 1407 (2011)

\bibitem{Margueritat}
J.~Margueritat, H.~Gehan, J.~Grand, G.~L\'evi, J.~Aubard, N.~F\'elidj,
  A.~Bouhelier, G.~Colas-Des-Francs, L.~Markey, C.~Marco De~Lucas, A.~Dereux,
  E.~Finot, ACS Nano \textbf{5}(3), 1630 (2011)

\bibitem{Halas12}
W.S. Chang, J.B. Lassiter, P.~Swanglap, H.~Sobhani, S.~Khatua, P.~Nordlander,
  N.J. Halas, S.~Link, Nano Lett. \textbf{12}(9), 4977 (2012)

\bibitem{Rogach}
T.K. Sau, A.L. Rogach (eds.), \emph{Complex-Shaped Metal Nanoparticles:
  Bottom-Up Syntheses and Applications} (Wiley-VCH Verlag GmbH \& Co, 2012)

\bibitem{Isabell99}
T.C. Isabell, P.E. Fischione, C.~O'Keefe, M.U. Guruz, V.P. Dravid, Microsc. Microanal. \textbf{5}, 126 (1999)

\bibitem{quidant10}
S.Graells, S A\'cimov\'ic, G. Volpe and R. Quidant; Plasmo. \textbf{5}, 135 (2010)

\bibitem{Schiffmann93}
K.I. Schiffmann, Nanotech. \textbf{4}(3), 163 (1993)

\bibitem{Antognozzi97}
M.~Antognozzi, A.~Sentimenti, U.~Valdr\`e, Microsc. Microanal. Microstruct.
  \textbf{8}(6), 355 (1997)

\bibitem{Johnson02}
S.~Johnson, D.~Hasko, K.~Teo, W.~Milne, H.~Ahmed, Microelectr. Eng.
  \textbf{61–62}(0), 665  (2002)

\bibitem{Djenizian03}
T.~Djenizian, L.~Santinacci, H.~Hildebrand, P.~Schmuki, Surf. Sci.
  \textbf{524}(1–3), 40  (2003)

\bibitem{Griffiths10}
A.J.V. Griffiths, T.~Walther, J. Phys.: Conference Series
  \textbf{241}(1), 012017 (2010)

\bibitem{toth09}
M.~Toth, C.J. Lobo, M.J. Lysaght, A.E. Vlad\'{a}r, M.T. Postek, J.
  Appl.  Phys. \textbf{106}(3), 034306 (2009)

\bibitem{Ebisawa}
S.~Horiuchi, T.~Hanada, M.~Ebisawa, in \emph{EMC 2008 14th European Microscopy
  Congress 1–5 September 2008, Aachen, Germany}, ed. by M.~Luysberg,
  K.~Tillmann, T.~Weirich (Springer Berlin Heidelberg, 2008), pp. 387--388

\bibitem{Vane12}
C.G. Morgan, R.~Vane, in \emph{Proc. SPIE}, \emph{Metrology, Inspection, and
  Process Control for Microlithography XXVI}, vol. 8324, ed. by A.Starikov
  (2012)

\bibitem{mock02}
J.~Mock, M.~Barbic, D.R. Smith, D.A. Schultz, S.~Schultz, J. Chem. Phys
  \textbf{116}, 6755 (2002)

\bibitem{bouhelier05jpcb}
A.~Bouhelier, R.~Bachelot, J.~Im, G.P. Wiederrecht, G.~Lerondel, S.~Kostcheev,
  P.~Royer, J. Phys. Chem. B \textbf{109}, 3195 (2005)

\bibitem{Kelly03}
K.L. Kelly, E.~Coronado, L.L. Zhao, G.C. Schatz, J. Phys.
  Chem. B \textbf{107}(3), 668 (2003)

\bibitem{mcfarland03}
A.D. McFarland, R.P. Van~Duyne, Nano Letters \textbf{3}(8), 1057 (2003)

\bibitem{Ogawa99}
T.~Ogawa, M.~Akabori, J.~Motohisa, T.~Fukui, Phys. B: Cond. Mat.
  \textbf{270}(3–4), 313 (1999)

\bibitem{storm05}
A.J. Storm, J.H. Chen, X.S. Ling, H.W. Zandbergen, C.~Dekker, J. Appl. Phys.
  \textbf{98}(1), 014307 (2005)

\bibitem{akiba10}
M.~Kotera, Y.~Akiba, Jap. J. Appl. Phys. \textbf{49}(6), 06GE08 (2010)

\bibitem{knowles07}
M.~Toth, C. J. Lobo, G. Hartigan, and W. R. Knowles, J. Appl. Phys. \textbf{101}, 054309 (2007)

\bibitem{Reimer}
L. Reimer, \emph{Scanning Electron Microscopy: Physics of Image Formation and Microanalysis} (Springer Series in Optical Sciences, 1998)

\bibitem{Borel76}
Ph. Buffat and J-P. Borel, Phys. Rev. A \textbf{13}(6), 2287 (1976)

\bibitem{Egerton}
R.F. Egerton, P. Li, and M. Malac, Micron \textbf{35}, 399 (2004)


\end{thebibliography}

\end{document}